\documentclass[preprint,aps]{revtex4}

\usepackage{graphicx}
\usepackage{dcolumn}
\usepackage{bm}

\begin{document}


\title{Group theory and octupolar order in URu$_2$Si$_2$}
\author{Annam\'aria Kiss}
 \altaffiliation{Department of Physics, Tohoku University, Sendai 980-8578, Japan}\author{Patrik Fazekas}
\affiliation{Research Institute for Solid State Physics and Optics,\\
Budapest 114, P.O.B. 49, H-1525 Hungary}

\date{\today}

\begin{abstract}
Recent experiments on URu$_2$Si$_2$ show that the low-pressure
hidden order is non-magnetic but it breaks time reversal
invariance. Restricting our attention to local order parameters of
$5f^2$ shells, we find that the best candidate for hidden order is
staggered order of either ${\cal T}_z^{\beta}$ or ${\cal T}_{xyz}$
octupoles. Group theoretical  arguments for the effect of
symmetry-lowering perturbations (magnetic field, mechanical
stress) predict behavior in good overall agreement with
observations. We illustrate our general arguments on the example
of a five-state crystal field model which differs in several
details from models discussed in the literature. The general
appearance of the mean field phase diagram agrees with the
experimental results. In particular, we find that a) at zero
magnetic field, there is a first-order phase boundary between
octupolar order and large-moment antiferromagnetism with
increasing hydrostatic pressure; b) arbitrarily weak uniaxial
pressure induces staggered magnetic moments in the octupolar
phase; and c) a new phase with different symmetry appears at large
magnetic fields.
\end{abstract}

\maketitle

\section{\label{sec:int}Introduction}

The nature of the so-called "hidden order" of the $T<T_0\approx
17{\rm K}$ phase of URu$_2$Si$_2$ has long been debated
\cite{harrison}. Taking strictly on-site local order parameters
only, U$^{4+}\rightarrow 5f^2$ shells can carry magnetic dipole,
electric quadrupole, magnetic octupole, and even higher multipole
order parameters. The full local symmetry is described by ${\cal
G}={\cal D}_{4h}{\otimes}{\cal G}_t$ where ${\cal D}_{4h}$ is the
tetragonal point group, and ${\cal G}_t=\{{\hat E},{\hat T}\}$ the
two-element group generated by the time reversal operator\cite{inui} ${\hat
T}$. The classification of the twelve most obvious
\cite{12} local order parameters is given in Table~\ref{tab:tetr}.
Being expressed as Stevens equivalents, all order parameters are
even under space inversion. The notation "$g$" and "$u$" in
Table~\ref{tab:tetr} refers to their parity under time reversal.

\begin{table}[ht]\caption{Symmetry classification of the local order parameters \cite{local}
for ${\bf B} =0$ (${\cal D}_{4h}$ notations \protect\cite{inui},
overline means symmetrization \cite{shiina97}).} \label{tab:tetr}
\centering
\begin{tabular}{|c|c||c|c|}
\hline \hline sym ($g$)& operator & sym ($u$) & operator\\[1mm]
\hline
$A_{1g} $ &  ${\cal E}$ & $A_{1u} $&$ {\overline{J_x J_y J_z (J_x^2 - J_y^2)}} $\\[1mm]
$A_{2g}$ & ${\overline{J_x J_y  (J_x^2 - J_y^2)}} $& $A_{2u}$ & $J_z$ \\
$B_{1g} $ & ${\cal O}_2^2$ &  $B_{1u} $ &  ${\cal T}_{xyz}= {\overline{J_x J_y J_z }}$\\
$B_{2g}$ & ${\cal O}_{xy}={\overline{J_x J_y }}$ & $B_{2u}$ &
${\cal T}^{\beta}_z={\overline{J_z(J_x^2-J_y^2)}}$\\
$E_g$ & $\{ {\cal O}_{xz}, {\cal O}_{yz} \}$ & $E_u$ & $\{ J_x, J_y \}$  \\
\hline\hline
\end{tabular}
\end{table}

About 30 years of work on one of the most intensively studied
$f$-electron systems has not brought clarification: the order is
still "hidden" \cite{harrison}. As we are going to describe,
theoretical progress has long been held up by the ambiguity of
experimental findings on apparently heterogeneous samples.
However, crucial recent experiments \cite{musrinpress,yokoyama}
allow to infer what the equilibrium properties of ideal samples of
URu$_2$Si$_2$ would be.

From the earliest neutron scattering experiments \cite{broholm},
the issue has been complicated by the observation of apparent
$f$-electron micromagnetism. Ascribing the magnetic moments to the
bulk of the sample, the observations indicated two-sublattice
${\bf Q}=(0,0,1)$  antiferromagnetism of U $5f$-shell moments of
$O(0.01\mu_{\rm B})$ directed along the tetragonal fourfold axis
$z$ in the low-$T$ ($T<T_0$) phase. Though the nominal value of
the ordered moment $m$ was two orders of magnitude lower than the
paramagnetic moment, this seemed to conform to the general idea
that micromagnetism is the canonical behavior of $f$-electron
systems on the borderline between the non-magnetic (heavy fermion)
Kondo state and RKKY magnetism \cite{colemanandrei}. According to
this view, URu$_2$Si$_2$ might have been put in the same class as
UPt$_3$ or CeAl$_3$ \cite{but}.

Many previous ideas about URu$_2$Si$_2$ were based on the
assumption that antiferromagnetism with micro-moments is a static
phenomenon, and it is an intrinsic feature of the $T<T_0$ phase.
Since the ordering of small moments could not account for the
large thermal anomalies at the 17K transition, it was assumed that
the staggered dipole moment $m$ is a secondary order parameter,
being induced by the primary ordering of an un-identified
full-amplitude order parameter $\psi$ (the hidden order). This
would require that antiferromagnetism has the same symmetry as the
hidden order, i.e., $\psi$ should break time reversal invariance
and share the spatial character of $m$ under the symmetry
classification according to the tetragonal point group \cite{notice} ${\cal
D}_{4h}$. With these assumptions, the Landau free
energy functional would contain a term $-m\psi$, generating $m\ne
0$ whenever the primary $\psi\ne 0$. This is a scenario which we
are going to discard, for the reasons given below, and further in
Sec.~\ref{sec:oct}.

The intimate connection between hidden order and micromagnetism
looked always somewhat suspicious because the variability
$(0.017-0.04)\mu_{\rm B}$ of the antiferromagnetic moment was too
large to be associated with nominally good-quality samples, and
because the onset of micromagnetism did not exactly coincide with
$T_0$. Susceptibility and NMR under pressure give an insight:
though the sample-averaged sublattice magnetization grows with
pressure, it seems to arise from the increase in the number of
magnetic sites, not from changing the magnetic moment at a given
site \cite{matsuda}. This points to the possibility that the
apparent micromagnetism is an attribute of heterogeneous samples,
and should be understood as ordinary antiferromagnetism of a small
($\sim 1\%$) volume fraction in samples which for some reason
always include a minority phase \cite{segreg}.

The argument was clinched by high-pressure $\mu$SR experiments:
hidden order is non-magnetic, and antiferromagnetism of at least
$O(0.1\mu_{\rm B})$ ionic moments appears at a first-order
transition at $p_{\rm tr}\approx 0.6{\rm GPa}$ (Ref.~\onlinecite{musrinpress}).
There are two thermodynamic phases, a non-magnetic phase with
$\langle\psi\rangle\ne 0$ and $\langle m\rangle= 0$, and the
antiferromagnetic phase with $\langle m\rangle\ne 0$ and $\langle
\psi\rangle= 0$. At ambient pressure, the magnetic ($\langle
m\rangle\ne 0$) phase is slightly less stable than the phase with
hidden order \cite{amitsuka}. However, large-amplitude
antiferromagnetism is stabilized at hydrostatic pressures
$p>0.6{\rm GPa}$ following a first order non-magnetic-to-magnetic
transition. In a range of low hydrostatic pressures, the nature of
the low-temperature phase remains the same as in ambient
conditions: $\psi\ne 0$ and $m=0$. The situation is, of course,
different if we apply fields which lower the symmetry of the
system: magnetic field ${\bf B}$, or uniaxial stress $\sigma$.

In the following Sections, we discuss the effect of uniaxial
stress, and of magnetic field, on the ordered phases of
URu$_2$Si$_2$. We will deduce that the low-pressure zero-field
order must be staggered octupolar order of either $B_{1u}$ or
$B_{2u}$ octupoles (Sec.~\ref{sec:oct}). The overall appearance of
the temperature - magnetic field phase diagram will be explained
(Sec.~\ref{sec:magn}). Finally, the general arguments will be
illustrated by the results obtained from a new crystal field model
(Sec.~\ref{sec:Xtal}).

\section{\label{sec:oct}Octupolar order}

In this Section, we argue that the experimental evidence presented
in Refs.~\onlinecite{musrinpress,yokoyama} unambiguously shows that the
"hidden order" of URu$_2$Si$_2$ is alternating octupolar order
with ${\bf Q}=(0,0,1)$. Here we restrict our attention to the
strictly local (on-site) order parameters \cite{tsuruta} listed in
Table~\ref{tab:tetr}. Two-site quadrupole--spin and three-site
spin--spin correlators could appear in the same symmetry class as
on-site octupoles \cite{agterberg}; the present argument does not
differentiate these cases.

Let us recall the hydrostatic pressure -- temperature phase
diagram obtained from high-pressure $\mu$SR experiments
\cite{musrinpress} (a phase diagram of the same shape results from
our mean field theory, see Fig.~\ref{fig:press}, left). Hidden
order $\langle\psi\rangle\ne 0$ is the attribute of the
low-pressure, low-temperature phase ($p<p_{\rm tr}$,
$T<T_{0}(p)$). Though all samples show some micromagnetism, it can
be safely concluded that this is an extrinsic effect and in a
perfect sample, hidden order should be non-magnetic.
Antiferromagnetism of at least $O(0.1\mu_{\rm B})$ ionic moments
appears at a first-order transition at $p_{\rm tr}\approx 0.6{\rm
GPa}$ . At $p<p_{\rm tr}$ the hidden order onset temperature
$T_0(p)$, and at $p>p_{\rm tr}$ the N\'eel temperature $T_{\rm
N}(p)$ are critical temperatures for the
hidden-order-to-paramagnetic, and the
antiferromagnetic-to-paramagnetic transitions, respectively; these
meet the first order phase boundary at a bicritical point. It
follows that $m$ and $\psi$ are of different symmetry, and the
Landau free energy cannot contain a term $m\psi$. The symmetry of
$\psi$ must be in any case different from $A_{2u}({\bf
Q}=(0,0,1))$.

Regarding the absence of magnetism as an established fact, we can
also exclude $E_u$ (magnetic moments perpendicular to the
tetragonal $z$-axis). The remaining choices for the order
parameter $\psi$ are: quadrupolar ($B_{1g}$, $B_{2g}$, or $E_g$),
octupolar ($B_{1u}$ or $B_{2u}$), hexadecapole ($A_{2g}$), or
triakontadipole ($A_{1u}$) (see Table~\ref{tab:tetr}). Quadrupole
and hexadecapole moments are time reversal invariant, while
octupoles and triakontadipoles change sign under time reversal.

An important recent experiment allows to decide the time reversal
character of the hidden order. Yokoyama et al \cite{yokoyama}
carried out magnetic neutron scattering measurement in the
presence of uniaxial stress applied to a single crystal sample
either along, or perpendicular to the tetragonal axis. Stress
$\sigma\parallel(001)$ does not produce significant change in
magnetic moments. However, for stress $\sigma\perp(001)$ the
staggered moment increases approximately linearly, reaching $\sim
0.25\mu_{\rm B}$ at $\sigma=0.4$GPa. In contrast to hydrostatic
pressure, no threshold value is needed to induce a magnetic
moment; it appears as soon as the stress $\sigma$ is finite.

Mechanical stress is time reversal invariant, thus it can produce
magnetic moments only from an underlying state which itself breaks
time reversal invariance. This limits the choice of hidden order
to $B_{1u}$ or $B_{2u}$ (octupolar), or $A_{1u}$
(triakontadipoles). We emphasize that the choice of octupolar
order is essentially different from the previously assumed
quadrupolar order \cite{santini1,santini2,ohkawa} which does not
break time reversal invariance. Additional evidence in favor of
the time reversal invariance breaking character of $\psi$ comes
from NMR measurements\cite{bernal}.

We will show that the properties of URu$_2$Si$_2$ can be described
well with the assumption of octupolar order. This would make
URu$_2$Si$_2$ the third well-argued case of primary octupolar
order in an $f$-electron system (the first two cases being NpO$_2$
\cite{npo2} and Ce$_{1-x}$La$_x$B$_6$ \cite{KK}). Within the
limits of our argument, either $B_{1u}$ or $B_{2u}$ would
reproduce the basic effect of stress-induced large-amplitude
antiferromagnetism. On the other hand, we rule out $A_{1u}$
triakontadipoles as order parameters.

First, we consider stress applied in the (100) direction.
$\sigma\parallel(100)$ lowers the symmetry to orthorombic ${\cal
D}_{2h}$ (see Appendix). Under ${\cal D}_{2h}$ , $A_{2u}^{\rm
tetr}\rightarrow B_{1u}^{\rm orth}$ and $B_{2u}^{\rm
tetr}\rightarrow B_{1u}^{\rm orth}$, so the order parameters
${\cal T}_z^{\beta}$ and $J_z$ become mixed
(Table~\ref{tab:orth}). A state with spontaneous ${\cal
T}_z^{\beta}$ octupolar order carries $J_z$ magnetic dipole
moments as well \cite{reverse}, accounting for the observations
\cite{yokoyama}.

An alternative way to derive this is by inspecting the relevant
terms of the Landau potential for the undistorted tetragonal phase
(the operators in the equations below have the meaning given in
Table~\ref{tab:tetr}). Choosing $B_{2u}$ octupolar order
parameter, consider the mixed third order invariant
\begin{eqnarray}
{\cal I}(A_{2u}{\otimes}B_{1g}{\otimes}B_{2u}) & = & c_1 J_z({\bf
0}) {\cal T}^{\beta}_z ({\bf Q}){\cal O}_2^2(-{\bf Q})
\nonumber\\
& & +c_2 J_z({\bf Q}) {\cal T}^{\beta}_z (-{\bf Q}){\cal
O}_2^2({\bf 0})\, . \label{eq:inv1}
\end{eqnarray}

 Generally, $c_1\ne 0$ and $c_2\ne 0$. For our
present purpose, the second term matters. A uniform stress
$\sigma\parallel(100)$ induces uniform (${\bf q}={\bf 0}$) ${\cal
O}_2^2$ quadrupole density \cite{strictly} which couples the
staggered (${\bf q}={\bf Q}$) $B_{2u}$ octupole order parameter to
the $J^z$ dipole density with ${\bf q}=-{\bf Q}$, i.e., the same
spatial modulation. Neutron scattering shows that stress-induced
antiferromagnetism has the same simple two-sublattice structure
with ${\bf Q}=(0,0,1)$ that was previously ascribed to
micromagnetism, thus the periodicity of the hidden octupolar order
must be the same.

\begin{figure}
\centering
\includegraphics[height=7cm,angle=270]{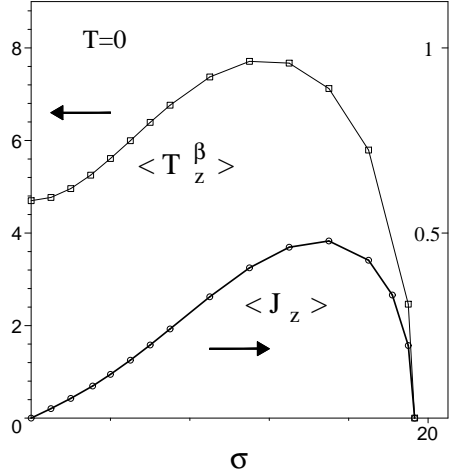}
\caption{Stress-induced magnetic moment in the
octupolar phase, based on the crystal field model described in
Sec.~\protect\ref{sec:Xtal}. Thick line: $\langle M_z\rangle$
staggered magnetization, thin line: $\langle {\cal
T}_z^{\beta}\rangle$ octupolar moment, as a function of the
uniaxial pressure $\sigma\parallel (100)$ ($\sigma$ in arbitrary
units).} \label{fig:stress}
\end{figure}

The actual stress-dependence of antiferromagnetic polarization
depends on microscopic details. Fig.~\ref{fig:stress} illustrates
the general behavior for a crystal field model which we describe
in detail later.

 Stress applied along the $z$-axis induces ${\cal
O}_2^0$ which transforms according to the identity representation
$A_{1g}$, thus it does not appear in the invariants, and it is not
predicted to induce magnetism. This is in qualitative accordance
with the observation that for $\sigma\parallel(001)$ the induced
moments are an order of magnitude smaller than for
$\sigma\parallel(100)$. We believe that the fact that these
moments are not exactly zero, is due to non-ideality of the sample,
as micromagnetism itself is.

The situation is less clear with varying the direction of stress
in the $\sigma\perp (001)$ plane. Experiments find that the
stress-induced antiferromagnetic moment is essentially the same
for $\sigma\|(110)$ as for $\sigma\|(100)$ \cite{yokoyama}. Taken
in itself, stress-induced antiferromagnetism would be as easy to
understand for $\sigma\|(110)$ as it was for $\sigma\|(100)$.
Namely, the invariant expansion of the Landau potential contains
also
\begin{eqnarray}
{\cal I}(A_{2u}{\otimes}B_{2g}{\otimes}B_{1u}) & = & c_3 J_z({\bf
0}) {\cal T}_{xyz} ({\bf Q}){\cal O}_{xy}(-{\bf Q})
\nonumber\\
& & +c_4 J_z({\bf Q}) {\cal T}_{xyz} (-{\bf Q}){\cal O}_{xy}({\bf
0})\, . \label{eq:inv2}
\end{eqnarray}
$\sigma\|(110)$ induces uniform ${\cal O}_{xy}$ quadrupolar
polarization. Assuming that the hidden (octupolar) order is ${\cal
T}_{xyz} (-{\bf Q})$, it is coupled to $J_z({\bf Q})$, the same
kind of antiferromagnetism as we found before. An alternative way
to arrive at the same result is by observing that $\sigma\|(110)$
lowers the symmetry to orthorombic, under which ${\cal T}_{xyz}$
and $J_z$ belong to the same irrep (Appendix,
Table~\ref{tab:orth}).

We have to emphasize, though, that assuming a homogeneous system,
either we have an explanation for the effect at $\sigma\|(100)$
(with ${\cal T}^{\beta}_z$ octupolar order), or for
$\sigma\|(110)$ (with ${\cal T}_{xyz}$ octupolar order), but not
for both. Under tetragonal symmetry, ${\cal T}^{\beta}_z$ and
${\cal T}_{xyz}$ belong to different irreps, and therefore these
orders cannot coexist. At the level of our present argument, the
problem cannot be resolved. We believe that it is not merely a
difficulty with our model but it points to a genuine feature of
URu$_2$Si$_2$. We speculate that the ${\cal T}^{\beta}_z$ and
${\cal T}_{xyz}$ orders are sufficiently near in energy, and so
samples tend to contain domains of both.

 We note that the $A_{1u}$
triakontadipole ${\overline{J_x J_y J_z (J_x^2 - J_y^2)}}$ (see
Table~\ref{tab:tetr}) would not give rise to stress-induced
magnetism and is therefore not a suitable choice as order
parameter.

It is worth pointing it out that our present scenario offers an
explanation why micromagnetism is always present. This may seem
paradoxical since if it were connected with a minority phase only,
it would be reasonable to expect that some preparation techniques
give single-phase samples, i.e., completely non-magnetic ones.
However, we ascribe antiferromagnetism also  to the polarization
of the primary octupolar phase in a stress field. It can be
assumed that the environment of impurities and crystal defects
always contains regions with the local stress oriented
perpendicularly to the tetragonal main axis, thus there is always
some local antiferromagnetism.

\section{\label{sec:magn}Magnetic field}

There have been extensive studies of the effect of an external
magnetic field on the phase diagram of URu$_2$Si$_2$
\cite{jaime,kimsuslovharr}. The system is relatively insensitive
to fields applied in the $x$--$y$ plane, while fields ${\bf
B}\parallel {\hat z}$ have substantial effect: hidden order can be
suppressed completely with $B_{{\rm cr},1}=34.7{\rm T}$. The phase
boundary in the $B$--$T$ plane is a critical line, thus hidden
order (or its suitable modification) breaks a symmetry also at
${\bf B}\ne {\bf 0}$. At somewhat higher fields, an ordered phase
appears in the field range $B_{{\rm cr},2}=35.8{\rm T}<B<B_{{\rm
cr},3}=38.8{\rm T}$. One possibility is that it is the re-entrance
of the $B<B_{{\rm cr},1}$ hidden order; however, we are going to
argue that the high-field order has different symmetry than the
low-field order.

The difference between the previously suggested quadrupolar order
\cite{santini1}, and our present suggestion of octupolar order, is
sharp at $B=0$ (Ref.~\onlinecite{alsoat}). However, a $B\ne 0$ magnetic field
mixes order parameters which are of different parity under time
reversal (Table~\ref{tab:tetrB}). The reason is that switching on
a field ${\bf B} \parallel {\hat z}$ lowers the point group
symmetry from ${\cal D}_{4h}{\otimes}{\cal G}_t$ to an 8-element
group isomorphic (but not identical) to ${\cal C}_{4v}$
(Ref.~\onlinecite{inui}).

Switching on a field ${\bf B} \parallel {\hat z}$, geometrical
symmetry is lowered from ${\cal D}_{4h}$ to ${\cal C}_{4h}$.
However, the relevant symmetry is not purely geometrical. Though
taken in itself, reflection in the $xz$ plane ${\hat
\sigma}_{v,x}$ is not a symmetry operation (it changes the sign of
the field), combining it with time reversal ${\hat T}$ gives the
symmetry operation ${\hat{\boldmath T}}{\hat \sigma}_{v,x}$. The
same holds for all vertical mirror planes, and ${\cal C}_2\perp
{\hat z}$ axes, thus the full symmetry group consists of eight
unitary and eight non-unitary symmetry operations \cite{inui}
\begin{equation}
{\cal G}(B_z) = {\cal C}_{4h} + {\hat{\boldmath T}}{\hat
\sigma}_{v,x} {\cal C}_{4h} \, .
\end{equation}
We may resort to a simpler description observing that
\begin{equation}
{\tilde{\cal G}} = {\cal C}_{4} + {\hat{\boldmath T}}{\hat
\sigma}_{v,x} {\cal C}_{4}
\end{equation}
is an important subgroup of ${\cal G}(B_z)$, and we can base a
symmetry classification on it. The multiplication table of
${\tilde{\cal G}}$ is the same as that of ${\cal C}_{4v}$, and
therefore the irreps can be given similar labels. It is in this
indirect sense that the symmetry  in the presence of a field ${\bf
B}
\parallel {\hat z}$ can be regarded as ${\cal C}_{4v}$ (a
convention used in Ref.~\onlinecite{shiina97}). The symmetry classification
of the local order parameters valid in ${\bf B} \parallel {\hat
z}$ is given in Table~\ref{tab:tetrB}. The results  make it
explicit that the magnetic field mixes dipoles with quadrupoles,
quadrupoles with certain octupoles, etc.

\begin{table}
\caption{Symmetry classification of the lowest rank
local order parameters for ${\bf B} \parallel {\hat z}$ (notations
as for ${\cal C}_{4v}$ \protect\cite{inui})} \label{tab:tetrB}
\centering
\begin{tabular}{|c||l|}
\hline \hline Symmetry & \ \ basis operators\\
\hline $A_1$ &  \ \ 1, $J_z$\\
$A_2$ & \ \ ${\overline{J_xJ_y(J_x^2-J_y^2)}}$,
${\overline{J_xJ_yJ_z(J_x^2-J_y^2)}}$\\
$B_1$ & \ \ ${\cal O}_2^2$, ${\cal T}^{\beta}_z$\\
$B_2$ & \ \ ${\cal O}_{xy}$, ${\cal T}_{xyz}$ \\
$E$ & \ \ $\{J_x, J_y\}$, $\{ {\cal O}_{xz}, {\cal O}_{yz}\}$\\
\hline\hline
\end{tabular}
\end{table}

\begin{figure}
\centering
\includegraphics[height=7cm,angle=0]{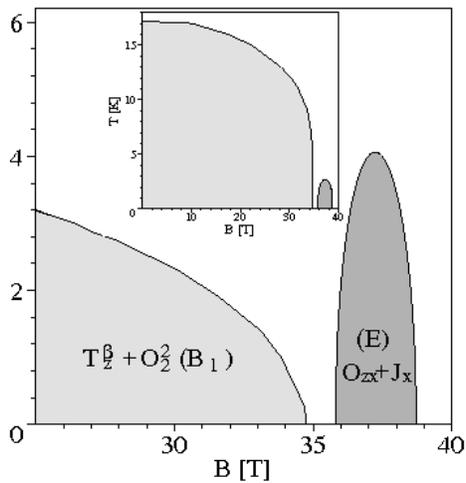}
\caption{The high-field part of the $T=0$ phase
diagram of the multipolar model ($B$ in units of T (Tesla)).
Vertical axis: $\langle {\cal T}^{\beta}_z\rangle$ for the
low--field phase, and $\langle {\cal O}_{zx}\rangle$ for the
high-field phase. The field-induced mixing of the order parameters
is shown within the shaded areas. The overall appearance of the
$T$-$B$ (inset, $T$ in units of K) phase diagram is very similar.
(The critical temperature of the $E$ phase is scaled up 3-fold).
\label{fig:phase}}\end{figure}

In a field ${\bf B} \parallel {\hat z}$, there can exist ordered
phases with four different local symmetries: $A_2$, $B_1$, $B_2$,
and $E$. The zero-field $B_{2u}$-type ${\cal T}^{\beta}_z$
octupolar order evolves into  the $B_{1}$-type ${\cal
T}^{\beta}_z$--${\cal O}_2^2$ mixed octupolar--quadrupolar order.
If the octupolar order is staggered, it mixes with similarly
staggered quadrupolar order: this follows from the first line of
Eqn.~(\ref{eq:inv1}) \cite{how}. The character of the low-field
phase is indicated in the ground state phase diagram in
Figure~\ref{fig:phase} (all numerical results are derived from a
crystal field model described in Sec.~\ref{sec:Xtal}, but the
validity of our general arguments is not restricted to that
particular model). The gradual suppression of octupolar order
under field applied in a high-symmetry direction is a well-known
phenomenon; a similar result was derived for $f^3$ ions in Ref.~\onlinecite{kf}. In our calculation, the octupolar phase is suppressed
at $B_{\rm cr,1} \approx 34.7{\rm T}$ (Figure~\ref{fig:phase}).

Although we are not familiar with experimental results for the
combined effect of hydrostatic pressure and magnetic field, it
should follow from our scheme that a critical surface is bounding
the phase with staggered $B_1$ octupolar--quadrupolar  order until
at sufficiently high pressures, the critical surface terminates by
a bicritical line. The high-pressure low-field phase has
alternating $J_z$ order like in the zero-field case. Hydrostatic
pressure does not change the symmetry of the system, but it can
change the numerical values of the coefficients in the expansion
of the Landau free energy in terms of invariants. Therefore,
generally speaking, we expect continuity with the results found
for $p=1{\rm atm}$ up to a threshold value of the pressure where a
first order transition to a phase with different symmetry may take
place.

Let us return to the case of $B\parallel {\hat z}$ field effects
at ambient pressure. The story of the gradual suppression of the
$B_1$ octupolar--quadrupolar phase is closed by itself; it might
have happened that there is only one ordered phase, surrounded on
all sides by the disordered ($A_1$) phase. However, as shown in
Table~\ref{tab:tetrB}, there are order parameters of different
($A_2$, $B_2$, $E$) symmetries; it depends on microscopic details
whether such orders are induced by sufficiently high fields. If
they are, they cannot coexist with $B_1$, so the corresponding
domains in the $B_z$--$T$ plane must be either disjoint from the
$B_1$, or, if they are pressed against each other, separated by a
first order phase boundary. Our Figure~\ref{fig:phase} illustrates
the former case, where an $E$ phase with mixed
quadrupolar--dipolar order (see Table~{\ref{tab:tetrB}) is
separated from the low-field $B_1$ phase by a narrow stretch of
the disordered phase. We observe that this model result bears a
close resemblance to the phase diagram determined by high-field
experiments \cite{jaime,kimsuslovharr}. We note that high-field
transport experiments add more phase boundaries to those
determined by static experiments \cite{harrison}. However, it is
often found that transport anomalies delineate regions which,
while showing interesting differences in the dominant conduction
mechanism, still belong to the same thermodynamic phase.
Therefore, we take the view that the boundaries shown in
Figure~\ref{fig:phase} are the most robust features of the phase
diagram, and the first step should be identifying the nature of
these.

An interesting possibility to recover a phase diagram of the same
shape would be to identify the high-field phase as the
"re-entrance" of the low-field $B_1$ phase. This possibility was
suggested in Ref.~\onlinecite{harrison}. However, our present model study
does not predict re-entrance.

\section{\label{sec:Xtal}Crystal field model}

The previous arguments were based on a symmetry classification of
the order parameters, and the conclusions are independent of the
details of the microscopic models that allow the emergence of the
ordered phases (in particular, zero-field octupolar order) which
we postulate. However, many physical properties (foremost the
temperature dependence of the susceptibility, but also the
specific heat) were fitted with apparent success by making
different assumptions about the nature of the zero-field hidden
order (either quadrupolar order of $5f^2$ shells
\cite{santini1,santini2,santiniphd}, or non-conventional density
waves \cite{dw}). Therefore it is important to show that our work
is not in conflict with findings for which alternative
explanations had been suggested but offers fits to the results of
standard measurements, which are at least comparable, and in some
cases better, than previous results.

\begin{figure}
\centering
\includegraphics[height=4cm,angle=270]{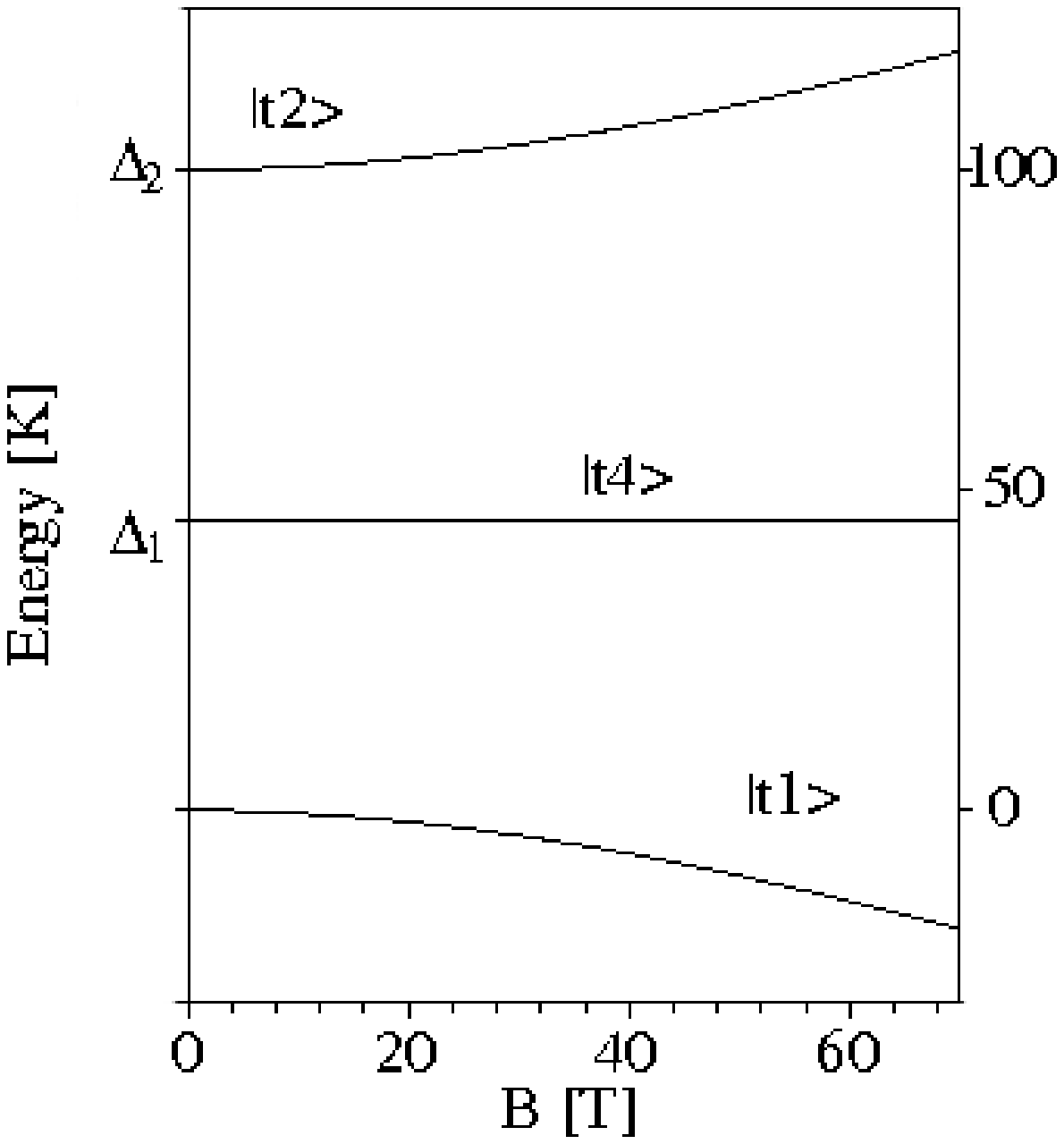}
\includegraphics[height=4.5cm,angle=270]{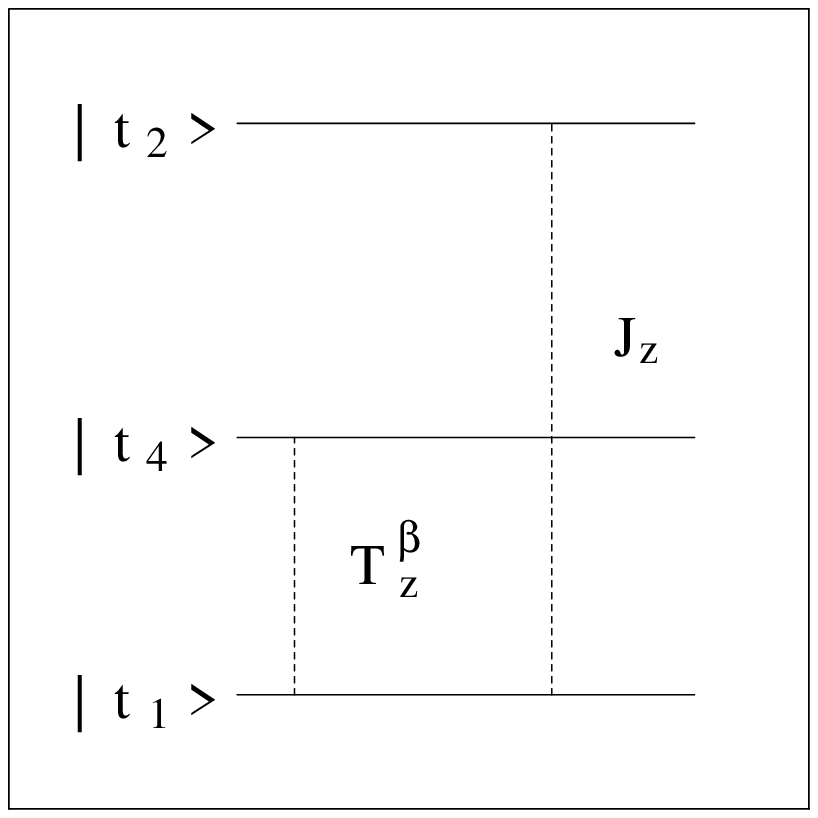}
\caption{In the low-field regime, the minimum model
consists of three singlets. {\sl Left}: The field-dependence of
the levels. {\sl Right}: relevant multipole matrix elements.
\label{fig:levelsa}}
\end{figure}

Here we assume that equilibrium phases other than the
superconducting phase, can be described in terms of localized
$f$-electrons, with stable $5f^2$ shells. We note that for many
other interesting $f$-electron systems (e.g., CeB$_6$ and
Pr-filled skutterudites) the localized-electron description of
multipolar ordering works well, in spite of the fact that for
certain physical quantities, consideration of the itinerant
aspects of $f$-electron behavior is indispensable.

 It is generally agreed \cite{butdoublet} that the crystal field
 ground state is a singlet, and that the salient feature of the
 level scheme is three low-lying singlets. Three singlets are
 sufficient to account for low-energy phenomena. It is found that
 further two states have to be taken into account to get a
 satisfactory fit for the susceptibility up to room temperature.
 We note that the nature of the high-field ordered phase has not
 been discussed in previous crystal field theories.

The backbone of our crystal field model is the inclusion of the
same three singlets as in the works of Santini and coworkers
\cite{santini1,santini2,santiniphd}, but in different order
(Table~\ref{tab:J4states}, Fig.~\ref{fig:levelsa} (left)). The
ground state is the $\left|t_1\right>$ singlet, and
$\left|t_2\right>$ an excited state lying at $\Delta_2=100$K.
$\left|t_1\right>$ and $\left|t_2\right>$ are connected by a
matrix element of $J^z$, as observed by neutron inelastic
scattering \cite{broholm}. The lower-lying singlet
$\left|t_4\right>$ is connected to the ground state by an
octupolar matrix element: this feature allows the existence of
induced octupolar order as the strongest instability of the system
(Fig.~\ref{fig:levelsa}, right). We remark that while other level
schemes may also allow octupolar order if one assumes a stronger
octupole--octupole interaction, our assumption seems most
economical.

\begin{table}
\caption{Tetragonal crystal field states used in the
model.}\label{tab:J4states} \centering
\begin{tabular}{|l|c|c|c|}
\hline \hline state & form & symmetry & energy[K]\\
\hline $\left|t_2\right>$ &
$1/\sqrt{2}(\left|4\right>-\left|-4\right>)$ &
$A_2$ & 100\\
$\left|d_{\pm}\right>$ & $a\left|\pm 3\right>-\sqrt{1-a^2}
\left|\mp 1\right>$ & $E$ & 51\\
$\left|t_4\right>$ & $1/\sqrt{2}(\left|2\right>-\left|-2\right>)$
&
$B_2$ & 45\\
$\left|t_1\right>$ &
$b(\left|4\right>+\left|-4\right>)+\sqrt{1-2b^2}
\left|0\right>$ & $A_1$ & 0\\
\hline\hline
\end{tabular}
\end{table}

\begin{figure}
\centering
\includegraphics[height=4cm,angle=270]{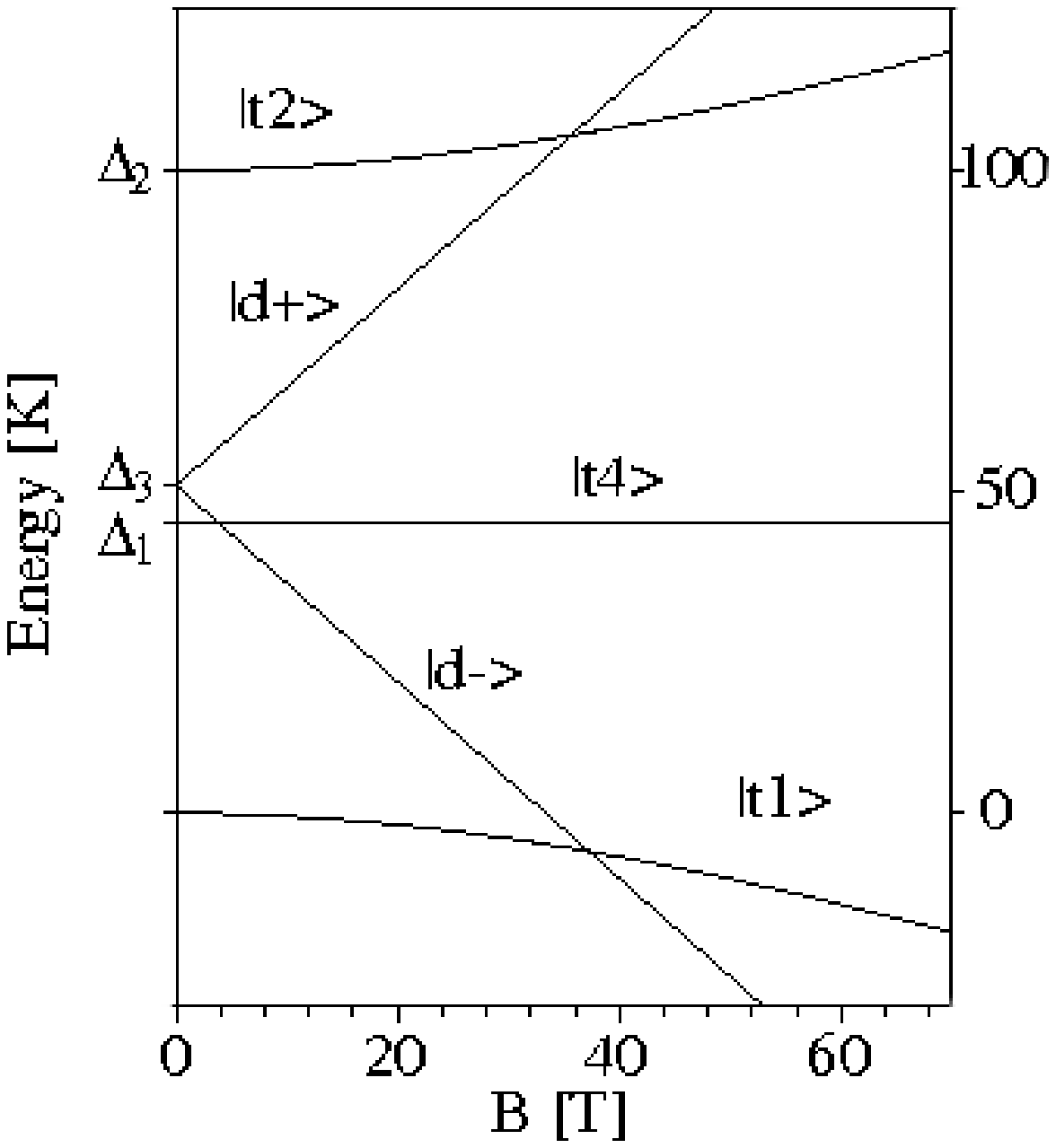}
\includegraphics[height=4.5cm,angle=270]{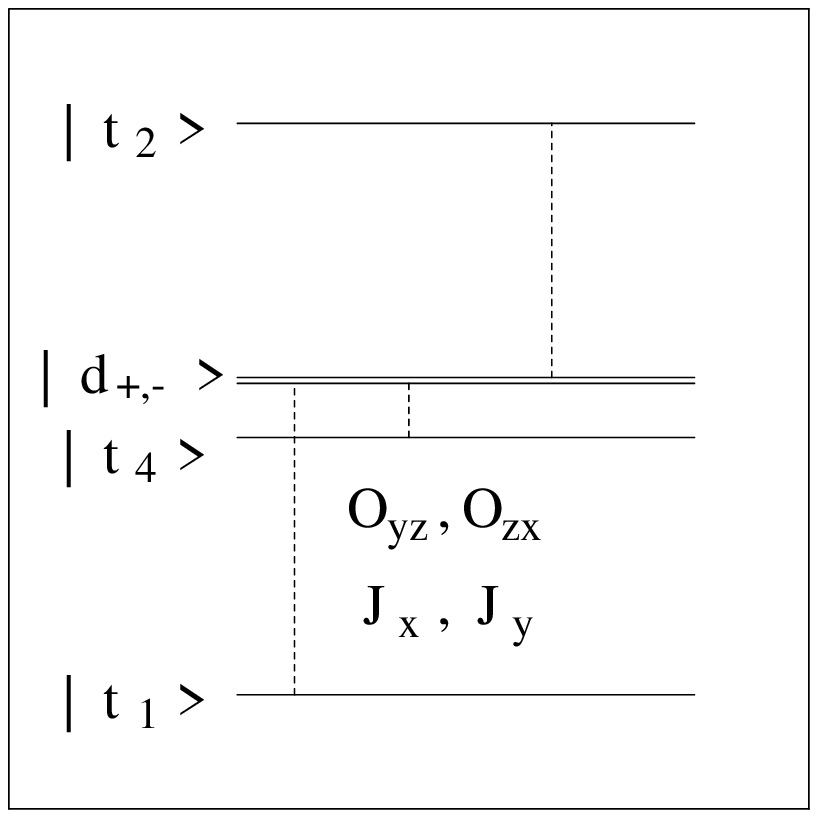}
\caption{{\sl Left}: The magnetic field dependence
of the single-ion levels in the extended five-state model used up
to high values of $T$ and ${\bf B}$. {\sl Right}: Additional
multipole matrix elements due to the addition of the doublet state
to the crystal field levels, which are relevant for the high-field
phase.\label{fig:levelsb}}
\end{figure}

Finally, as in previous schemes, at least two further states are
needed to fit magnetization data up to 300K. We found it useful to
insert one of the doublets ($\left|d_{\pm}\right>$). This is an
alternative to models with five singlets
\cite{santini1,santiniphd}. As we are going to see, fits to
standard macroscopic measurements are no worse in our scheme than
in previous ones. However, our scheme has the advantage that it
accounts for the high-field observations. We show the
field-dependence (${\bf B}\parallel {\hat z}$) of the crystal
field levels in Fig.~\ref{fig:levelsb}. The salient feature is the
crossing of the (mildly field-dependent) singlet ground state with
one of the levels derived from the splitting of the doublet at a
field strength lying between the critical fields $B_{\rm cr,2}$
and $B_{\rm cr,3}$. The crossing levels are connected by matrix
elements of $E$ operators (Table~\ref{tab:tetrB}). Consequently,
we find a high-field  $E$ phase where $\{J_x, J_y\}$-type
transverse dipolar order is mixed with $\{ {\cal O}_{zx}, {\cal
O}_{yz} \}$-type quadrupolar order (see Figure~\ref{fig:phase}).

Commenting on differences between our crystal field scheme
(Table~\ref{tab:J4states} where we use $a=0.98$, $b=0.22$) and
previously suggested ones, we note that unambiguous determination
is very difficult even if an intense experimental effort is
undertaken, as in the recent case of Pr-filled skutterudites. By
and large we agree with Nagano and Igarashi \cite{nagano}, who
argue that the crystal field potential of URu$_2$Si$_2$ is not
known in sufficient detail yet. We complied with constraints which
appear well-founded, as e.g. the neutron scattering evidence by
Broholm et al \cite{broholm}, but otherwise we adjusted the model
to get low-field octupolar order for which we found
model-independent arguments. Level positions were adjusted to get
good overall agreement with observations but we did not attempt to
fine-tune the model, neither did we check for alternative schemes
with less straightforward parametrization.

We use the mean field decoupled hamiltonian
\begin{eqnarray}
{\cal H}_{\rm MF}  =  \Delta_1 |t_4\rangle \langle t_4| + \Delta_2
|t_2\rangle \langle t_2| + \Delta_3 \sum_{\alpha
=+,-}|d_{\alpha}\rangle \langle d_{\alpha}|
 \nonumber\\
 -g{\mu}_{\rm B}B J_z
+ \lambda_{\rm oct}\left<{\cal T}_{z}^{\beta}\right>{\cal
T}_{z}^{\beta} -\lambda_{\rm quad}\left<{\cal O}_{zx}\right>{\cal
O}_{zx} \label{eq:mean}
\end{eqnarray}
where $g=4/5$, and the octupolar mean field coupling constant
$\lambda_{\rm oct}$ is meant to include the effective coordination
number; similarly for the quadrupolar coupling constant
$\lambda_{\rm quad}$. We assume alternating octupolar order, and
uniform ${\cal O}_{zx}$ order; the result would be the same if the
high-field quadrupolar order is also alternating. We do not
introduce independent ${\cal O}_2^2$ or $\{J_x$, $J_y\}$
couplings, nevertheless $\langle {\cal O}_2^2\rangle \ne 0$ in the
$B_1$ phase, and $\langle J_x\rangle\ne 0$ in the $E$ phase.

At $B=0$, the only non-vanishing octupolar matrix element is
$C=\langle t_1|{\cal T}_{z}^{\beta}|t_4\rangle \approx 8.8$.
Octupolar order is driven by  the large $C$: assuming
$\lambda_{\rm oct}=0.336$K we get the critical temperature
$T_{0}(B=0)=17.2$K for ${\cal T}_{z}^{\beta}$-type
antiferro-octupolar order. Using a similar estimate, we find
$\lambda_{\rm oc}^{\rm Np}\approx 0.2$K for NpO$_2$ which orders
at 25K (Ref.~\onlinecite{npo2}). The order-of-magnitude correspondence between
two documented cases of octupolar order shows that our present
estimate of the octupolar coupling strength is not unreasonable.

(\ref{eq:mean}) was solved for all temperatures and fields $B_z$.
We find that the octupolar phase is bounded by a critical line of
familiar shape (Fig.~\ref{fig:phase}, inset), which has its
maximum $T_0=17.2$K at $B_z=0$, and drops to zero at $B_{{\rm
cr},1}=34.7$T. The transition remains second order through-out; we
did not hit upon a tricritical point, though we are aware of no
reason of why it should not have appeared.

Similarly, the ground-state amplitude of the octupolar order is a
monotonically decreasing function of $B_z$ (Fig.~\ref{fig:phase}).

The restricted model with three singlets (Fig.~\ref{fig:levelsa})
offers two basic choices. In the absence of symmetry-lowering
fields, the Landau expansion of the free energy in terms of the
order parameters is
\begin{eqnarray}
{\cal F}&=&\alpha_{O}(T,p)\langle
T_{z}^{\beta}\rangle^2+\beta_{O}(T,p)\langle
T_{z}^{\beta}\rangle^4+\alpha_{M}(T,p)\langle
J_{z}\rangle^2\nonumber\\
&&+\beta_{M}(T,p)\langle J_{z}\rangle^4+...
\label{eq:landau0}\end{eqnarray} Note that because of the
tetragonal symmetry, the free energy expansion does not contain
the term $\langle T_{z}^{\beta}\rangle\langle J_{z}\rangle$. It
follows that the possible ordered phases can be (A) $\langle
T_{z}^{\beta}\rangle\ne 0$ and $\langle J_{z}\rangle=0$ or (B)
$\langle T_{z}^{\beta}\rangle=0$ and $\langle J_{z}\rangle\ne 0$.
This is in agreement with the experimental finding
\cite{musrinpress} that (A) is the low-pressure phase, and (B) is
the high-pressure phase, and they are separated by a first-order
boundary.

This canonical case  (qualitatively agreeing with the schematic
phase diagram shown in Ref.~\onlinecite{musrinpress}) is illustrated in
Fig.~\ref{fig:press} (left). It was derived from
Eqn.~(\ref{eq:mean}) using an ad hoc model assumption about the
pressure dependence of the crystal field splittings $\Delta_1$ and
$\Delta_2$ (Fig.~\ref{fig:press}, right) \cite{pressure}. The
shape of the phase boundaries, and in particular the slope of the
first order line, could be fine-tuned by adjusting the pressure
dependence of the crystal field parameters, but the overall
appearance of the phase diagram: two critical lines meeting at a
bicritical point, which is also the end-point of a first-order
boundary, is generic.

\begin{figure}
\centering
\includegraphics[height=4.25cm,angle=270]{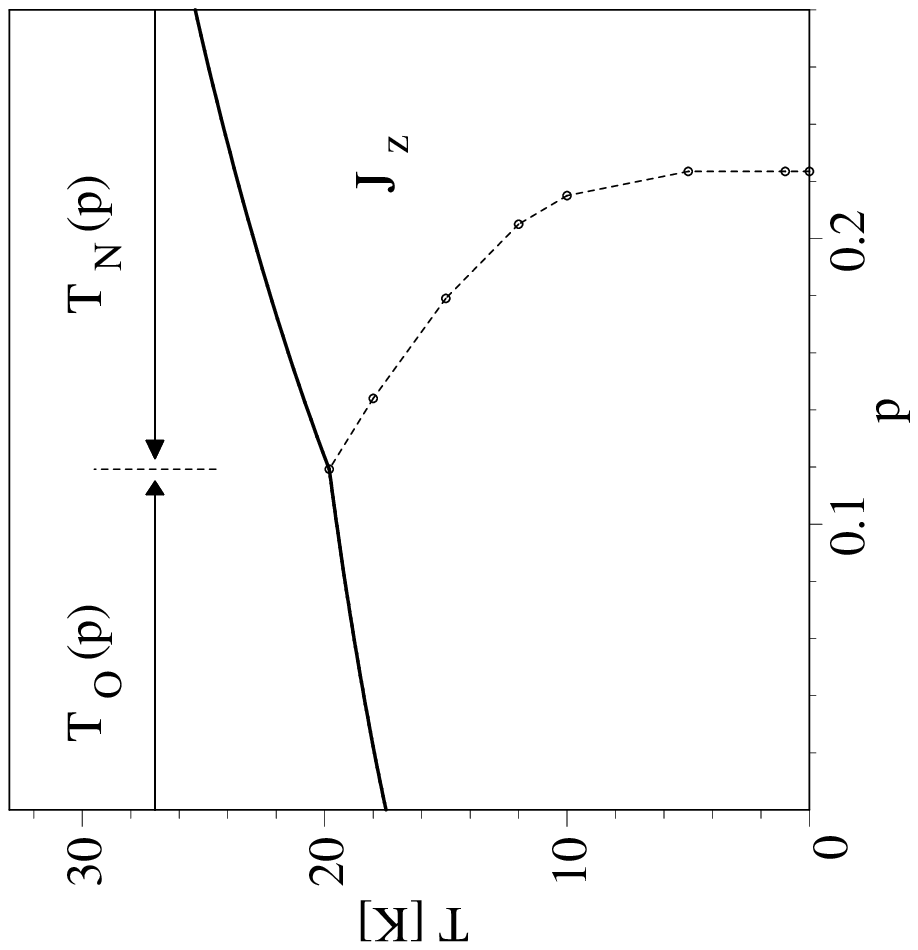}
\includegraphics[height=4.25cm,angle=270]{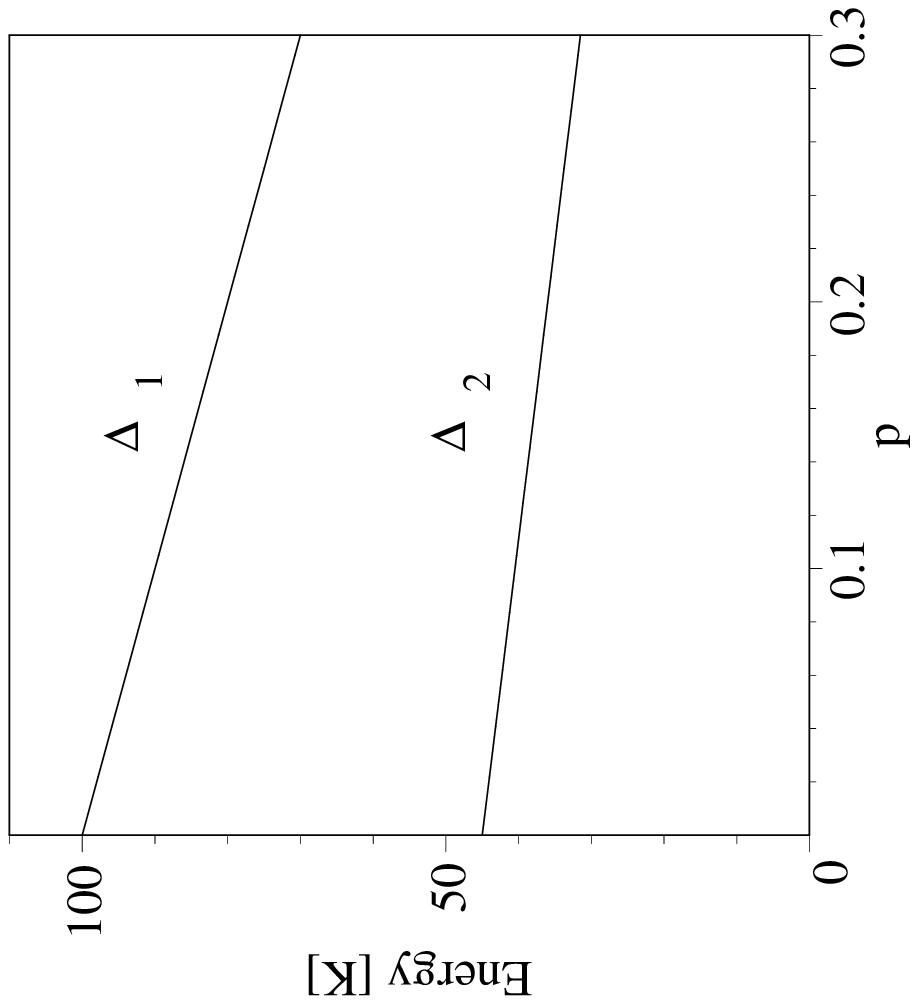}
\caption{{\sl Left}: The pressure dependence of the
critical temperature of the octupolar and the dipolar
antiferromagnetic phases ($T_0(p)$ and $T_N(p)$, respectively).
 The first-order boundary between the two ordered phases is an interpolation
through the calculated points. {\sl Right}: model assumption about
the pressure dependence of the crystal field splittings $\Delta_1$
and $\Delta_2$.}\label{fig:press}
\end{figure}

\begin{figure}
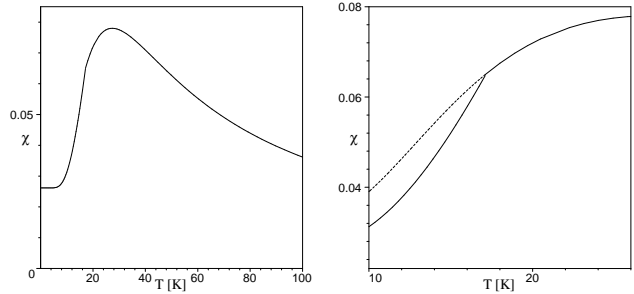

\centering
\includegraphics[height=4.25cm]{urusi_Fig_6_left.ps}
\includegraphics[height=4.25cm]{urusi_Fig_6_right.ps}
\caption{Linear susceptibility per site (in
$\mu_{\rm B}/{\rm T}$) on extended temperature scale (left), and
in the vicinity of the octupolar transition (right). The dashed
line gives the single-ion result. \label{fig:chi1}}
\end{figure}

The stress dependence of the induced antiferromagnetic moment
(Fig.~\ref{fig:stress}) was determined in a similar
 calculation, adding the term $-\sigma{\cal O}_2^2$
to the Landau potential, and solving the self-consistency
equations for $\langle J_z\rangle$ and $\langle{\cal
T}_z^{\beta}\rangle$. $\sigma$ in this calculation has the
character of uniaxial stress, but an additional set of
experimental data would be needed to determine its absolute scale.

 Next, we consider the results of some
standard low-field measurements. This was not the primary purpose
of our work but rather serves as a check. The quadrupolar model
\cite{santini1} obtained a reasonably good fit for the temperature
dependence of the linear and non-linear susceptibility in a range
of temperature, and we have to prove that our model yields a
comparably good description on a completely different microscopic
basis.

The octupolar transition shows up as a discontinuity of the linear
susceptibility (Fig.~\ref{fig:chi1}). This character of the
$\chi_1$ anomaly is expected from general arguments \cite{sakak}.
While the low-temperature behavior, including the regime around
$T_0$,  is satisfactorily described by the three-state model
(Fig.~\ref{fig:levelsa}), fitting the susceptibility up to room
temperature (Fig.~\ref{fig:chi1}, left) requires the five-state
model (Fig.~\ref{fig:levelsb}). One of the hallmarks of the
hidden-order transition of URu$_2$Si$_2$ is the strong jump of the
non-linear susceptibility $\chi_3$ (Ref.~\onlinecite{ramirez}). The shape of
the calculated anomaly (Fig.~\ref{fig:chi3}) corresponds
rather well to the experimental result.

\begin{figure}
\centering
\includegraphics[height=6cm,angle=270]{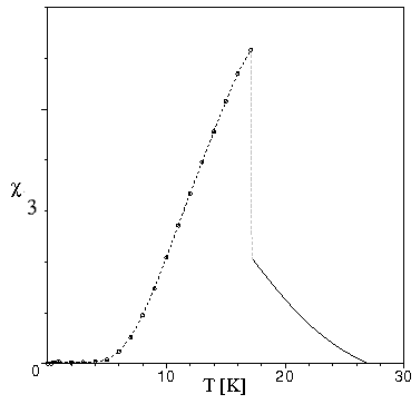}
\caption{The temperature dependence of
the nonlinear susceptibility $\chi_3$ in the vicinity of the
octupolar transition. The dashed line is an interpolation through
the calculated points\label{fig:chi3}}
\end{figure}

\begin{figure}
\centering
\includegraphics[height=6.5cm,angle=270]{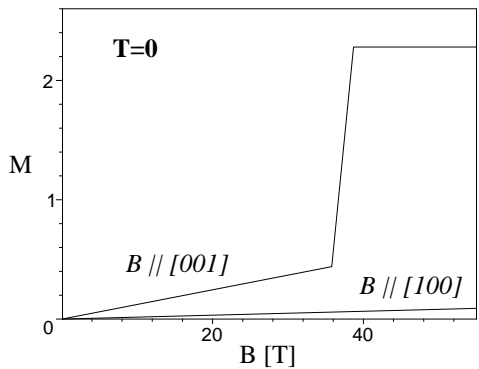}
\caption{The magnetization curve at
$T=0$ ($M$ in units of    $\mu_{\rm B}$). \label{fig:chi32}}
\end{figure}

Next we discuss the high-field behavior at $B>B_{{\rm cr},1}$, and
interpret the disjoint high-field phase observed in experiments
\cite{jaime,kimsuslovharr} as a mixed quadrupolar--dipolar phase
(the $E$ phase in Fig.~\ref{fig:phase}). We exploit the field
dependence of the ionic levels in the five-level model
(Table~\ref{tab:J4states}, Fig.~\ref{fig:levelsb}). The single-ion
levels $t_1$ and $d_-$ would cross at $B_{\rm cross}=37.3$T. Since
$|t_1\rangle$ and $|d_-\rangle$ are connected by $E$ operators
including ${\cal O}_{zx}$ (see Fig.~\ref{fig:levelsb}, right), a
range of fields centered on $B_{\rm cross}$ is certain to favour
$\{ {\cal O}_{zx},{\cal O}_{yz} \}$ quadrupolar order,  and
simultaneous $\{ J_x, J_y \}$ dipolar order. We chose a weak
quadrupolar interaction $\lambda_{\rm quad}=0.054$K in
Eqn.~(\ref{eq:mean}); this gives quadrupolar order between the
critical fields $B_{{\rm cr},2}=35.8$T and $B_{{\rm cr},3}=38.8$T.
The amplitude of quadrupolar order is not small
(Fig.~\ref{fig:phase}) but the ordering temperature is low ($\sim
1$K) because the coupling is weak. The $E$ phase shows up as the
steep part of the magnetization curve in Fig.~\ref{fig:chi32}. For $\lambda_{\rm quad}=0$ we would have a jump-like
metamagnetic transition at  $B=B_{\rm cross}$.

We are aware of an unsatisfactory feature of the calculated
magnetization curve. Though it is clear that our theory involves
three critical fields: $B_{{\rm cr},1}$, $B_{{\rm cr},2}$, and
$B_{{\rm cr},3}$, at the lowest of these the anomaly is so weak
that it does not show up on the scale of Fig.~\ref{fig:chi32}. We get a single-step metamagnetic transition distributed
over the width of the high-field quadrupolar phase. The overall
height of the step is right, but we do not recover the three-step
structure of the transition observed by Sugiyama et al
\cite{sugiyama}.

\section{\label{sec:conc}Discussion and Conclusion}

There have been many attempts to explain the non-superconducting
phases of URu$_2$Si$_2$. Though the behavior of $f$-electrons in
this system certainly has itinerant aspects, or perhaps
URu$_2$Si$_2$ is on the verge of a localized-to-itinerant
transition, arguing on the basis of a simple localized electron
model can lead to useful results. Namely, crystal field theory
conforms to a general symmetry classification of the equilibrium
phases, which is expected to apply to a wider range of models,
including suitably defined Kondo lattice, or Anderson lattice,
models. Our main interest lies in cross-effects like the mixing of
order parameters in the presence of external magnetic field, or
mechanical stress. Our conclusions rely on symmetry reasoning, and
only numerical details depend on the choice of the crystal field
model which we use to illustrate the general arguments.

The identification of the low-pressure, low-temperature hidden
order of URu$_2$Si$_2$ is of basic interest. Starting from the
high-temperature tetragonal phase, a symmetry-breaking transition
can lead to an ordered phase with the following choices for the
local order parameter:  $A_{2u}$ and $E_u$ dipoles, $B_{1g}$,
$B_{2g}$, and $E_g$ quadrupoles, $B_{1u}$, and $B_{2u}$ octupoles,
a $A_{2g}$ hexadecapole, and an $A_{1u}$ triakontadipole
\cite{local}.

It was always clear that the primary order parameter of
URu$_2$Si$_2$ cannot be dipolar. The possibility of quadrupolar
ordering has been extensively discussed \cite{santini1}. Higher
multipoles have been mentioned in a general context
\cite{agterberg,bourdarot}, but have not been studied in detail.

 A recent $\mu$SR study \cite{musrinpress} finds
that the symmetry of hidden order is different from $A_{2u}({\bf
Q})$ which is the symmetry of the high-pressure antiferromagnetic
phase (the same structure was ascribed to the supposed
"micromagnetism" of URu$_2$Si$_2$, which is now understood to be
extrinsic). The present experimental status is that the intrinsic
low-pressure behavior of URu$_2$Si$_2$ is purely non-magnetic.
Furthermore, a number of recent experiments proves that the hidden
order breaks time reversal invariance, so it cannot be quadrupolar
\cite{bernal,bourdarot,yokoyama}. In particular, Yokoyama et al
\cite{yokoyama} found that uniaxial stress (which is time reversal
invariant) induces large-amplitude antiferromagnetism, which
breaks time reversal invariance. It is clear that stress must have
acted on a medium which itself was non-invariant under time
reversal: it must have been the octupolar phase \cite{triak}.

We emphasize that stress-induced antiferromagnetism arises only if
the stress is uniaxial, and perpendicular to the (001) direction.
For tetragonal symmetry, octupoles ($B_{1u}$ and $B_{2u}$), and
dipoles ($A_{2u}$ and $E_u$) are of different symmetry and
therefore they do not mix. It follows that hydrostatic pressure
cannot induce antiferromagnetism unless the pressure is high
enough to lead to a completely different (purely dipolar) phase
via a first-order phase transition. This was found in
Ref.~\cite{musrinpress}. In contrast, uniaxial pressure
perpendicular to the tetragonal main axis lowers the symmetry to
orthorombic, allowing the mixing of dipoles and octupoles.

We postulated that the hidden order is ${\cal T}^{\beta}_z$
staggered octupolar order (Sec.~\ref{sec:oct}). Uniaxial pressure
$\sigma\|(100)$ leads to the appearance of $J_z$ dipolar order of
the same periodicity. The model works the same way if we postulate
${\cal T}_{xyz}$ staggered octupole order, in which case a stress
$\sigma\|(110)$ gives $J_z$ antiferromagnetism. Since the
octupoles ${\cal T}^{\beta}_z$ and ${\cal T}_{xyz}$ belong to
different one-dimensional irreps of the tetragonal symmetry
($B_{2u}$ and $B_{1u}$, respectively), in our theory a homogeneous
system can show only one of the stress-induced effects. We
hypothesized that the observed near-equivalence of the stress
effect in (100) and (110) directions \cite{yokoyama} reflects the
presence of both kinds of order in a multi-domain structure.

The same assumption about ${\cal T}^{\beta}_z$ octupolar order
explains the behavior in applied magnetic field
(Sec.~\ref{sec:magn}). A field ${\bf B}\|(001)$ mixes ${\cal
T}^{\beta}_z$ octupoles with ${\cal O}_2^2$ quadrupoles. Symmetry
breaking is well-defined in the presence of magnetic field, and
the transition to hidden order (now a mixed octupolar--quadrupolar
order) remains second order up to a critical field $B_{{\rm
cr},1}$ where $T_0(B)\to 0$.

We illustrated the symmetry arguments on the example of a
crystal-field model (Sec.~\ref{sec:Xtal}). The model has two
versions: low-energy phenomena can be described by using three
low-lying singlets, while for high energies (or fields, or
temperatures) we need five states (the previous three singlets
plus a doublet). The three singlets are the same as in Santini's
work, but their sequence was chosen to give an octupolar matrix
element between the ground state and the first excited state. The
presence of the doublet level is not essential at low fields (and
low temperatures) but it splits in a magnetic field ${\bf
B}\|(001)$, and for a range of high fields, even weak quadrupolar
coupling can give quadrupolar order which competes with the
low-field octupolar order. We argued that the high-field order
observed between 35T and 38T \cite{jaime,kimsuslovharr} is of
quadrupolar nature, with a symmetry different from that of the
low-field order.

To conclude, we presented arguments showing that octupolar order
of either $B_{2u}$ or $B_{1u}$ symmetry is the zero-field "hidden
order" of URu$_2$Si$_2$ at ambient conditions. We limited the
discussion to strictly on-site order parameters in a localized
electron model with stable $5f^2$ valence. However, within this
restriction our scenario is compatible with the present knowledge
about the phase diagram in the temperature--pressure--field space.
The time reversal invariance breaking nature of the order is
manifest in the effect that uniaxial pressure applied in certain
directions can induce large-amplitude antiferromagnetism.

\begin{acknowledgments}
The authors are greatly indebted to Yoshio
Kuramoto for inspiring discussions, valuable advice, and
continuing encouragement. We were supported by the Hungarian
National Grants OTKA T038162, T037451, and TS040878. A.K. is
recipient of a COE fellowship at the Tohoku University (Sendai),
and acknowledges support by the Hungarian--Japanese Joint Project
"Competition and Frustration in Multipolar Ordering Phenomena".
\end{acknowledgments}

\appendix*

\section{Orthorombic symmetry}

Quadrupolar moments couple to external stress. For instance,
applying stress $\sigma \|(100)$, ${\cal O}_2^2$ quadrupolar
moments are induced, while $\sigma \|(110)$ induces ${\cal
O}_{xy}$. At the same time, the application of uniaxial stress
lowers the symmetry from the tetragonal ${\cal D}_{4h}$ to one of
its subgroups, changing the symmetry classification of all order
parameters. This effect is described below.

When we apply uniaxial pressure in direction $(100)$, the previous
$C_4$, $S_4$, $C_2^{''}$ and $\sigma_v$ cease to be symmetry
operations and the residual symmetry is described by the group
${\cal D}_{2h}$. The corresponding classification of the order
parameters is given in Table~\ref{tab:orth}. We observe that under
the new symmetry, the ${\cal T}^{\beta}_z$ octupolar and the $J_z$
dipolar moments mix with each other, and this means that if the
system possesses spontaneous staggered ${\cal T}^{\beta}_z$
octupolar order, applying $\sigma \|(100)$ stress induces
staggered $J_z$ dipolar moments.

\begin{table}[ht]\caption{Symmetry classification of the
local order parameters for $\sigma \|(100)$ ({\sl left part}) and for
$\sigma \|(110)$ ({\sl right part}). The subscripts  $g$
and $u$ mean the even and odd under time reversal.
$H_z^{\alpha}={\overline{J_x J_y (J_x^2 - J_y^2)}}$ is one of the
nine hexadecapoles.} \label{tab:orth} \centering
\begin{tabular}{c|c}
\hline \hline$D_{2h}$ &\\[1mm]
\hline
$A_{1g} $ &  ${\cal O}_2^2$\\
$B_{1g}$ &   ${\cal O}_{xy}$, $H_z^{\alpha}$\\
$B_{2g} $ & ${\cal O}_{zx}$\\
$B_{3g}$ & ${\cal O}_{yz}$\\
\hline
$A_{1u} $ &  ${\cal T}_{xyz}$\\
$B_{1u}$ & ${\cal T}^{\beta}_z$, $J_z$\\
$B_{2u} $ & $J_x$\\
$B_{3u}$ &  $J_x$\\
\hline\hline
\end{tabular}
\hspace*{20mm}
\begin{tabular}{c|c}
\hline \hline$D_{2h}$ &\\[1mm]
\hline
$A_{1g} $ &  ${\cal O}_{xy}$\\
$B_{1g}$ &  ${\cal O}_2^2$ , $H_z^{\alpha}$\\
$B_{2g} $ & ${\cal O}_{yz}-{\cal O}_{zx}$\\
$B_{3g}$ & ${\cal O}_{yz}+{\cal O}_{zx}$\\
\hline
$A_{1u} $ & ${\cal T}^{\beta}_z$ \\
$B_{1u}$ & ${\cal T}_{xyz}$, $J_z$\\
$B_{2u} $ & $J_x+J_y$\\
$B_{3u}$ &  $J_x-J_y$\\
\hline\hline
\end{tabular}
\end{table}

When the uniaxial pressure is applied in $(110)$ directions, it
induces ${\cal O}_{xy}$ quadrupoles. Now $C_4$, $S_4$, $C_2^{'}$
and $\sigma_d$ have to be omitted from the symmetry group which is
again the ${\cal D}_{2h}$ point group, only comprised of different
elements than in the $\sigma \|(100)$ case. For the present
$\sigma \|(110)$ case, the symmetry classification of the order
parameters is shown in Table~\ref{tab:orth}. Now the ${\cal
T}_{xyz}$ octupolar moment mixes with the $J_z$ dipolar moment (if
${\cal T}_{xyz}$ is staggered, so is $J_z$).

When we apply uniaxial pressure along the tetragonal main axis
$(001)$, there is no symmetry reduction, the original $D_{4h}$
symmetry classification of the order parameters
(Table~\ref{tab:tetr}) remains valid. Neither the ${\cal T}_{xyz}$
nor the ${\cal T}^{\beta}_z$ octupolar moments can induce $J_z$
magnetic moments, since they all correspond to different
irreducible representations of the ${\cal D}_{4h}$ point group. An
analogous statement holds for the staggered moments.


\begin{thebibliography}{99}

\bibitem{harrison} N. Harrison, K.H. Kim, M. Jaime, and J.A. Mydosh,
  Physica B {\bf 346}-{\bf 347}, 92 (2004), and references therein.

\bibitem{inui} T. Inui, Y. Tanabe, and Y. Onodera: {\sl Group Theory and Its
Applications in Physics}, Springer Series in Solid-State Sciences
{\bf 78}, Springer-Verlag, Berlin (1990).

\bibitem{12} These are the twelve lowest-rank multipoles which are needed to get an
operator for each of the entries in Table~\ref{tab:tetr}; they
were enumerated, e.g., in Refs.~\onlinecite{santini2,santiniphd}. A
sufficiently large local Hilbert space would support also order
parameters which do not appear in Table~\ref{tab:tetr}.  Though
this holds  for our five-state-model described in
Sec.~\ref{sec:Xtal} (it supports 24 non-trivial local order
parameters, i.e., 12 more than we listed), we do not discuss them
because in our interpretation they would correspond to
higher-energy processes.

\bibitem{local} We call local order parameters strictly on-site multipole moments.
Pair and three-site correlators can also be defined in terms of
localized $5f^2$ models \cite{agterberg} but we do not consider
them.

\bibitem{shiina97} We follow closely the method developed by R. Shiina,
H. Shiba, and P. Thalmeier: J. Phys. Soc. Japan {\bf 66}, 1741
(1997). See also Ref.~\onlinecite{kf}.

\bibitem{musrinpress} A. Amato, M.J. Graf, A. de Visser, H. Amitsuka, D. Andreica,
and A. Schenck: J. Phys.: Condens. Matter {\bf 16},  S4403 (2004);
H. Amitsuka et al: Physica B {\bf 326}, 418 (2003).

\bibitem{yokoyama}
M. Yokoyama, H. Amitsuka, K. Tenya, K. Watanabe, S. Kawarazaki, H. Yoshizawa, J. A. Mydosh, cond-mat/0311199 (2003)

\bibitem{broholm} C. Broholm, J. K. Kjems, W. J. L. Buyers, P. Matthews, T. T. M. Palstra, A. A. Menovsky, J. A. Mydosh, Phys. Rev. Lett. {\bf 58}, 1467 (1987).

\bibitem{colemanandrei} P. Coleman and N. Andrei: J. Phys.:
Condens. Matter {\bf 1}, 4057 (1989).

\bibitem{but} We note, however, that the status of micromagnetism as an intrinsic
static phenomenon is doubtful in at least several other systems as
well (see Ref.~\onlinecite{musrinpress}, and A. Schenck and G. Solt: J.
Phys.: Condens. Matter {\bf 16},  S4639 (2004)).

\bibitem{notice} Notice that this cannot be realized within
Table~\ref{tab:tetr}. Dipoles are either $A_{2u}$, or $E_u$; their
partners in the same row of Table~\ref{tab:tetr} are
time-reversal-even ($E_g$ quadrupoles and $A_{2g}$ hexadecapoles,
respectively). However, in principle suitable $A_{2u}$, or $E_u$
partners to $J_z$, $\{J_x,J_y\}$ may be found by considering
operators spanned by higher-lying states of the local Hilbert
space, or allowing non-local order parameters.

\bibitem{matsuda} K. Matsuda, Y. Kohori, T. Kohora, K. Kuwahara, H. Amitsuka, Phys. Rev. Lett. {\bf 87}, 087203 (2001).

\bibitem{segreg} H. Amitsuka and M. Yokoyama, Physica B {\bf 329}-{\bf
333}, 452 (2003).

\bibitem{amitsuka} H. Amitsuka, K. Tenya, M. Yokoyama, A. Schenck, D. Andreica, F. N. Gygax, A. Amato, Y. Miyako, Ying Kai Huang, J. A. Mydosh, Physica B {\bf 326}, 418
(2003).

\bibitem{tsuruta} In a valence fluctuating $f^2$--$f^1$ model,
it is possible to ascribe octupolar moments solely to conduction
electrons (see A. Tsuruta, A. Kobayashi, T. Matsuura and Y.
Kuroda: J. Phys. Soc. Japan {\bf 69}, 663 (2000); J. Phys. Chem.
Solids {\bf 62}, 301 (2001)). We do not consider this possibility.
In any case, Tsuruta et al consider the hidden order as
quadrupolar, and associate octupolar order with a second phase
transition at 15K, for which there is no compelling experimental
evidence.

\bibitem{agterberg} D.F. Agterberg and M.B. Walker, Phys. Rev. B {\bf 50}, 563
  (1994).

\bibitem{santini1} P. Santini, Phys. Rev. B {\bf 57} 5191 (1998).

\bibitem{santini2}
P. Santini and G. Amoretti, Phys. Rev. Lett. {\bf 73} 1027 (1994).

\bibitem{ohkawa} F.J. Ohkawa and H. Shimizu: J. Phys.: Condens. Matter {\bf 11},
L519 (1991)

\bibitem{bernal} O.O. Bernal, C. Rodrigues, A. Martinez, H. G. Lukefahr, D. E. MacLaughlin, A. A. Menovsky, J. A. Mydosh, Phys. Rev. Lett. {\bf 87},
196402 (2001).


\bibitem{npo2} P. Santini and G. Amoretti, Phys. Rev. Lett. {\bf 85}, 2188
(2000), and references therein.

\bibitem{KK} H. Kusunose and Y. Kuramoto: J. Phys. Soc. Japan {\bf 70},
1751 (2001); K. Kubo and Y. Kuramoto: J. Phys. Soc. Japan {\bf
73}, 216 (2004).

\bibitem{reverse} The reverse statement is also true: if
the primary order is $J_z$, in the presence of $\sigma\| (100)$
stress the system acquires $T^{\beta}_z$ octupolar moment. This
may be relevant at high uniaxial pressures.

\bibitem{strictly} Strictly speaking, $\sigma\|(100)$ induces the
quadrupole $3{\cal O}_2^2-{\cal O}_2^0\propto 3x^2-r^2$, but
${\cal O}_2^0$ transforms like identity, so for our purposes,
${\cal O}_2^2$ is induced.

\bibitem{jaime} M. Jaime, K. H. Kim, G. Jorge, S. McCall, J. A. Mydosh, Phys. Rev. Lett. {\bf 89}, 287201 (2002).

\bibitem{kimsuslovharr} K.H. Kim, N. Harrison, M. Jaime, G. S. Boebinger, J. A. Mydosh, Phys. Rev. Lett. {\bf 91}, 256401
  (2003);  A. Suslov, J. B. Ketterson, D. G. Hinks, D. F. Agterberg, Bimal K, Sarma, Phys. Rev. B {\bf 68} 020406 (2003);
N. Harrison, M. Jaime, J. A. Mydosh, Phys. Rev. Lett. {\bf 90}, 096402 (2003).

\bibitem{alsoat} The mixed octupolar--quadrupolar character is the
same as in the work of Santini \cite{santini1}. However, the $B\to
0$ limit is pure octupole order in our present model, while it is
pure quadrupole order in Santini's model.

\bibitem{how} Field effects can be treated on the basis of the
zero-field expansion of the Landau potential, understanding that
in each term, one of the factors {\bf J} has to be replaced by
{\bf B} (see Ref.~\onlinecite{kf}).


\bibitem{kf} A. Kiss and P. Fazekas: Phys. Rev. B {\bf 68}, 174425 (2003).

\bibitem{santiniphd} P. Santini, {\sl Thesis} (Lausanne, 1994).

\bibitem{dw}  P. Chandra, P. Coleman, J.A. Mydosh and V. Tripathi,
Nature (London) {\bf 417}, 831 (2002); A. Virosztek, K. Maki, and
B. D\'ora, Int. J. Mod. Phys. {\bf 16}, 1667 (2002).

\bibitem{butdoublet} With the exception of the suggestion of a
doublet ground state in the crystal field model of
Ref.~\onlinecite{ohkawa}. The valence fluctuation model of Tsuruta et al
\cite{tsuruta} assumes that the $f^2$ component of the ground
state is a doublet.

\bibitem{nagano} T. Nagano and J. Igarashi: cond-mat/0409735.

\bibitem{pressure} The parametrization shown in
Fig.~\ref{fig:press} (right) defines the pressure scale of the
schematic phase diagram. We did not attempt to find the
relationship to the actual pressure scale.

\bibitem{sakak} T. Sakakibara, T. Tayama, K. Kitami, M. Yokoyama, K. Tenya, 
H. Amitsuka, D. Aoki, Y. Onuki, Z. kletowski, T. Matsumra, and T. Suzuki: 
J. Phys. Soc. Japan {\bf 69}, Suppl. A, 25 (2000).

\bibitem{ramirez} A.P. Ramirez, P. Coleman, P. Chandra, E. Br\" uk, A. A. Menovsky, Z. Fisk, E. Bucher, Phys. Rev. Lett. {\bf 68}, (1992) 2680.

\bibitem{sugiyama} K. Sugiyama, H. Fuke, K. Kindo, K. Shimohata, A. A. Menovsky, J. A. Mydosh, M. Date, J. Phys.
Soc. Japan {\bf 59}, 3331 (1990)

\bibitem{bourdarot} The possibility of octupolar order was discussed in a
  different context by 
F. Bourdarot, B. Fak, K. Habicht, and K.
Prokes: J. Magn. Magn. Mater. {\bf 272}-{\bf 276}, e31 (2004), in a work 
describing the 
results of inelastic neutron scattering experiments carried out at high
fields. In this work, the conclusion about octupolar order is based on the
assumption of $m$ and $\psi$ sharing the same symmetry, thus it is different
from our present argument. We thank F. Bourdarot and B. Fak for enlightening
correspondence on this issue. 

\bibitem{triak} Triakontadipole order is
time-reversal-invariance-breaking, but it does not allow the
interpretation of the uniaxial stress effect.


\end{thebibliography}

\end{document}